\documentclass[aps,prl,preprint,groupedaddress,showpacs]{revtex4-1}

\usepackage{graphicx}
\usepackage{bm}

\begin{document}

\title{Universal description of three two-component fermions }

\author{O.~I.~Kartavtsev}\email{oik@nusun.jinr.ru}
\author{A.~V.~Malykh}\email{maw@theor.jinr.ru}
\affiliation{Joint Institute for Nuclear Research, Dubna, 141980, Russia}

\date{\today}

\begin{abstract} 
A quantum mechanical three-body problem for two identical fermions of mass $m$ 
and a distinct particle of mass $m_1$ in the universal limit of zero-range 
two-body interaction is studied. 
For the unambiguous formulation of the problem in the interval 
$\mu_r < m/m_1 \le \mu_c$ ($\mu_r \approx 8.619$ and $\mu_c \approx 13.607$) 
an additional parameter $b$ determining the wave function near 
the triple-collision point is introduced; thus, a one-parameter family 
of self-adjoint Hamiltonians is defined. 
The dependence of the bound-state energies on $m/m_1$ and $b$ in the sector 
of angular momentum and parity $L^P = 1^-$ is calculated and analysed with 
the aid of a simple model. 
\end{abstract}


\maketitle


Low-energy dynamics of few two-species particles has attracted much attention 
as a basic quantum problem that is closely related to the investigations of 
ultra-cold binary quantum 
gases~\cite{Ospelkaus06,Fratini12,Iskin06,Levinsen09,Mathy11,Alzetto12,Jag14}. 
The principal problem is the investigation of few two-species fermions, 
in particular, the present Letter is aimed to study two identical fermions 
of mass $m$ interacting with a distinct particle of mass $m_1$. 
Since the few-body properties become independent of the particular form of 
the short-range two-body interaction in the low-energy limit, the universal 
description is obtained by using the contact or zero-range interaction 
defined by a single parameter, the two-body scattering length $a$. 
As a consequence, one expects that for the properly chosen units 
the few-body properties depend on a single non-trivial parameter, the mass 
ratio $m/m_1$. 

Significant advance was made in~\cite{Efimov73}, where it was demonstrated 
that for $m/m_1 > \mu_c$ ($\mu_c \approx 13.607$), similarly to 
the three-boson case, the problem of three two-species fermions is ambiguously 
defined in the limit of zero-range interaction. 
For the correct formulation, an additional parameter is needed to define 
the oscillating wave function near the triple-collision point. 
By setting this parameter, one comes to the Efimov spectrum, which contains 
an infinite number of bound states whose binding energies tend to infinity 
and the ratio of subsequent energies tends to a constant. 

For $m/m_1 \le \mu_c$, one of the important results was the analytic 
zero-energy solution, which reveals the two-hump structure in the low-energy 
three-body recombination rate dependence on $m/m_1$~\cite{Petrov03}. 
The three-body energy spectrum and the scattering cross sections 
for $L^P = 1^-$ were studied in~\cite{Kartavtsev07}, where two bound states 
were disclosed for $m/m_1$ increasing to $\mu_c$. 
The conclusions of~\cite{Kartavtsev07} were confirmed 
in~\cite{Endo11,Helfrich11} by solving the momentum-space integral equations. 
The formation of the three-body clusters should affect the properties of 
fermionic mixtures, in particular, it indicates effective attraction between 
a diatomic molecule and a light particle in the $p$-wave state, which persists 
even if the three-body system is unbound. 
In this respect, a role of the $p$-wave $2 + 1$ scattering was discussed 
in~\cite{Levinsen09,Levinsen11,Mathy11,Alzetto12} and the molecule-atom 
$p$-wave attraction in $^{40}$K--$^6$Li mixture was detected in~\cite{Jag14}. 
Furthermore, the dynamics of the ultra-cold gas consisting of three-body 
clusters was investigated~\cite{Naidon16,Endo15}. 
Another application to the many-body dynamics was the calculation of the third 
virial coefficient in the unitary limit $a \to \infty$~\cite{Daily12,Gao15}. 

In spite of progress, it is still necessary to correctly formulate 
the three-body problem for two-species fermions with zero-range two-body 
interaction in the mass-ratio interval $m/m_1 \le \mu_c$, as indicated in both 
physical~\cite{Nishida08,Safavi-Naini13,Kartavtsev14} and 
mathematical~\cite{Minlos12,Minlos14a,Correggi12,Correggi15} works. 
In this respect, the basic question is the unambiguous definition of 
the wave function in the vicinity of the triple-collision point. 
In this Letter, an additional three-body parameter $b$ is introduced 
to formulate the three-body problem for $\mu_r < m/m_1 \le \mu_c$ 
($\mu_r\approx 8.619$) that corresponds to the construction of a one-parameter 
family of self-adjoint Hamiltonians. 
Within the framework of this formulation, comprehensive analytic and 
numerical study of the three-body bound states is performed. 
Due to the permutational symmetry of fermions, the states of unit total 
angular momentum $L$ and negative parity $P$ are of most interest at low 
energy; for this reason, the $L^P = 1^-$ sector is considered in this Letter. 


The Hamiltonian in the centre-of-mass frame is the six-dimensional 
kinetic-energy operator $H_0 = -\Delta_{{\mathbf x}} - \Delta_{{\mathbf y}}$, 
where ${\mathbf x}$ and ${\mathbf y}$ are the scaled Jacobi coordinates 
and the units $\hbar = 2m/(1 + m/m_1) = 1$ are used. 
The two-body interaction is defined by the boundary condition for the wave 
function $\Psi$ imposed on two hyper-planes corresponding to the zero distance 
$r$ between either fermion and a distinct particle, 
$\displaystyle \lim_{r \rightarrow 0}\frac{\partial \ln (r\Psi)}{\partial r} 
= - \frac{1}{a}$. 
As the wave function is antisymmetric under permutation of fermions, a single 
condition in one pair of interacting particles is needed~\cite{Kartavtsev07}. 

The formal construction of the Hamiltonian does not obviously provide 
an unambiguous definition of the three-body problem; in particular, one should 
inspect the solution at the intersection of hyper-planes (triple-collision 
point). 
To analyse the wave function, correctly define the three-body problem, and 
calculate the binding energies, it is suitable to expand the wave function 
$\displaystyle\Psi = \rho^{-5/2}\sum_{n=1}^{\infty}f_n(\rho)\Phi_n(\rho, \Omega)$ 
into a set of eigenfunctions $\Phi_n(\rho, \Omega )$ of the auxiliary problem 
on a hyper-sphere at fixed $\rho $, where $\rho = \sqrt{x^2 + y^2}$ is 
a hyper-radius and $\Omega $ denotes a set of hyper-angular 
variables~\cite{Kartavtsev07}. 
This leads to an infinite set of coupled hyper-radial equations (HREs), 
\begin{equation}
\left[\frac{d^2}{d \rho^2} - \frac{\gamma_n^2(\rho) - 1/4}{\rho^2} + 
E \right] f_n(\rho) - \sum_{m = 1}^{\infty}\left[P_{nm}(\rho) - Q_{nm}(\rho)
\frac{d}{d\rho} - \frac{d}{d\rho}Q_{nm}(\rho) \right] f_m(\rho) = 0 \ ,
\label{system1}
\end{equation}
where the eigenvalues of the auxiliary problem $\gamma^2_n(\rho)$ are 
different branches of the multi-valued function defined for $L^P = 1^-$ by 
\begin{equation}
\frac{\rho }{a} \cos\gamma\frac{\pi}{2} = 
\frac{1 - \gamma^2}{\gamma}
\sin\gamma\frac{\pi}{2} - 2\frac{\cos\omega\gamma}{\sin2\omega}
 + \frac{\sin\omega\gamma}{\gamma\sin^2\omega} 
\label{transeq}
\end{equation}
and the notation $\sin\omega = 1/(1 + m_1/m)$ is used. 
The coupling terms $Q_{nm}(\rho)$ and $P_{nm}(\rho)$ are expressed in 
the analytical form via $\gamma^2_n(\rho)$ and their 
derivatives~\cite{Kartavtsev99,Kartavtsev06,Kartavtsev07}. 


Since both eigenfunctions $\Phi_n(\rho, \Omega )$ of the auxiliary problem and 
the coupling terms $Q_{nm}(\rho)$ and $P_{nm}(\rho)$ are regular, the wave 
function $\Psi $ for $\rho \to 0$ is basically determined by one of the channel 
functions $f_n(\rho)$, which corresponds to the least singular term 
$(\gamma_n^2 - 1/4)/\rho^2$ in the system of HREs~(\ref{system1}), i.~e., to 
the smallest $\gamma_n^2 $. 
For the sake of brevity, the channel index denoting the smallest eigenvalue, 
$\gamma^2(\rho)$, and the corresponding channel function, $f(\rho)$, will be 
omitted. 
To determine the channel function $f(\rho)$ up to the leading-order terms for 
$\rho \to 0$, one should retain in HRE the singular part 
$(\gamma^2 - 1/4)/\rho^2 + q/\rho$, where the notations 
$\gamma \equiv \gamma(0)$ and 
$q = \displaystyle\left[ \frac{d\gamma^2(\rho)}{d\rho} \right]_{\rho = 0}$ are 
introduced for brevity~\cite{footnote1}. 
Generally, $f(\rho) = C_+ \varphi_+ (\rho) + C_- \varphi_-(\rho)$ is a linear 
combination of two independent solutions, which up to the leading-order terms 
for $\rho \to 0$ are given by $\varphi_\pm(\rho) = \rho^{1/2 \pm \gamma} 
\left( 1 + \displaystyle\frac{q \rho}{1 \pm 2 \gamma}\right) $, except 
$\gamma = 0, 1/2$ when the expressions for $\varphi_\pm(\rho)$ contain 
logarithmic terms. 

Consider firstly $\gamma^2 \geq 1$ 
($m/m_1 \leq \mu_r \approx 8.619$)~\cite{footnote1}, in which case 
$\varphi_-(\rho)$ is not square-integrable when $\rho \to 0$ and should be 
excluded, i.~e., $C_- = 0$. 
Thus, one should satisfy the simple condition 
$f(\rho) \longrightarrow_{\hspace{-.6cm} \stackrel{\phantom{\chi \to}}{\rho \to 0}} 0$, 
in other words, the requirement of square integrability of $\Psi$ is 
sufficient. 
Conversely, if $\gamma^2 < 1$ ($m/m_1 > \mu_r$), both $\varphi_+(\rho )$ and 
$\varphi_-(\rho )$ are square-integrable and an additional boundary condition 
is needed if $\rho \to 0$. 
One should further distinguish the case $\gamma^2 < 0 $ 
($m/m_1 > \mu_c \approx 13.607$)~\cite{footnote1}, then $\varphi_\pm(\rho)$ 
oscillate and a standard method to lift ambiguity of the solution is 
to specify the constant $C_-/C_+$, which must satisfy $|C_-/C_+| = 1$ 
to provide self-adjointness of the Hamiltonian. 
Thus, one comes to the family of Hamiltonians depending on a single parameter 
(the phase of $C_-/C_+$) with the well-known Efimov spectrum of bound 
states~\cite{Efimov73}. 

The aim of this Letter is the unambiguous formulation of the problem for 
$1 > \gamma^2 \geq 0$ ($\mu_r < m/m_1 \leq \mu_c$), which requires one 
defining the boundary condition for $\rho \to 0$. 
Again, a standard method is to specify $C_-/C_+$, which should be real-valued 
to provide self-adjointness of the Hamiltonian. 
It is convenient to define the length $-\infty < b < \infty$ by
$-C_-/C_+ = \pm |b|^{2\gamma} \equiv b |b|^{2\gamma - 1} $, i.~e., $\pm $ refers 
to the sign of $b$. 
The boundary condition is straightforwardly written as 
\begin{equation}
\label{as_gam121}
\displaystyle f(\rho)
\longrightarrow_{\hspace{-.6cm} \stackrel{\phantom{\chi \to}}{\rho \to 0}} 
\rho^{1/2 + \gamma} \mp |b|^{2 \gamma} \rho^{1/2 - \gamma} 
\left[1 + q \rho /(1 - 2 \gamma )\right] \  
\end{equation}
except for $\gamma = 1/2$ ($m/m_1 = \mu_e \approx 12.313$~\cite{footnote1}). 
The last term $\sim q$ can be optionally omitted if $1/2 > \gamma > 0$ 
($\mu_e < m/m_1 < \mu_c$) and should be retained if $1 > \gamma > 1/2$ 
($\mu_r < m/m_1 < \mu_e$), when it exceeds the first term 
$ \rho^{1/2 + \gamma}$. 
If $\gamma = 0$ ($m/m_1 = \mu_c$), one finds the boundary condition either 
from Eq.~(\ref{as_gam121}) in the limit $\gamma \to 0$ or directly from  
$\varphi_+ \sim \sqrt{\rho}$ and $\varphi_- \sim \sqrt{\rho}\log (\rho)$, 
\begin{equation}
\label{as_gam0}
\displaystyle f(\rho) 
\longrightarrow_{\hspace{-.6cm} \stackrel{\phantom{\chi \to}}{\rho \to 0}} 
\rho^{1/2} \log (\rho /b)\ , 
\end{equation}
where only $b > 0$ is allowed. 
In the specific case of $\gamma = 1/2$ ($m/m_1 = \mu_e$) one can take 
$\varphi_+ \sim \rho$ and $\varphi_- \sim 1 + q \rho \log \rho$, which gives 
the boundary condition 
\begin{equation}
\label{as_gam12}
\displaystyle f(\rho) 
\longrightarrow_{\hspace{-.6cm} \stackrel{\phantom{\chi \to}}{\rho \to 0}} 
\rho - b (1 + q \rho \log \rho) \ . 
\end{equation}
As all other channel functions $f_n(\rho)$ tend to zero faster than 
$f(\rho)$ at $\rho \to 0$, it is sufficient to impose the conditions 
$f_n(0) = 0$ for complete formulation. 
For rigorous formulation the boundary condition should be imposed on 
the wave function $\Psi$, in particular, for $\mu_e < m/m_1 < \mu_c$ 
($1/2 > \gamma > 0$) it follows from~(\ref{as_gam121}) that 
\begin{equation} 
\label{psibc_gam120}
\displaystyle\lim_{\rho \to 0} \left( \rho^{1 - 2\gamma } 
\frac{d \log(\rho^{2 + \gamma}\Psi)}{d\rho } \pm \frac{2\gamma}{|b|^{2\gamma}}
\right) = 0 \ ; 
\end{equation}
however, for $\mu_r < m/m_1 < \mu_e$ ($1 > \gamma > 1/2 $) the boundary 
condition for $\Psi$ becomes cumbersome. 
In addition, the boundary condition is conveniently written in terms of 
the channel function $f(\rho)$ and its derivative~\cite{footnote1}. 
One should emphasise that the boundary condition~(\ref{as_gam12}) does not 
follow from~(\ref{as_gam121}) in the limit $\gamma \to 1/2$ and there is no 
continuous correspondence of the parameter $b$ defined for $\gamma = 1/2$ 
by Eq.~(\ref{as_gam12}) and that defined by Eq.~(\ref{as_gam121}). 
It is suitable to consider separately the dependence on $b$ for 
$m/m_1 = \mu_e$ ($\gamma = 1/2$)~\cite{footnote1}.  

The boundary condition imposed when $\rho \to 0$ is equivalent to including 
a zero-range three-body potential, while $b$ admits an interpretation as 
the generalised scattering length. 
This potential represents either the effect of intersection of the two-body 
potentials or the true three-body force. 
This interpretation can be illustrated by the connection of $b$ with 
the parameters of a particular potential, whose range is allowed to shrink 
to zero~\cite{footnote1}.  


The solution is simple in the limit $|a| \to \infty$ due to decoupling of 
HREs~(\ref{system1}), since the eigenvalues of Eq.~(\ref{transeq}) are 
constants $\gamma_n^2(0)$ independent of $\rho$ and the coupling terms 
$Q_{nm}(\rho)$ and $P_{nm}(\rho)$ vanish. 
Picking out one HRE with the smallest $\gamma_n^2(0) \equiv \gamma^2$ from 
the uncoupled system of HREs~(\ref{system1}) one finds for $b > 0$ that 
there is one bound state whose energy 
$ E = -4b^{-2}\left[-\Gamma(\gamma )/\Gamma(-\gamma )\right]^{1/\gamma }$ 
and eigenfunction $f(\rho) = \rho^{1/2} K_{\gamma }(\sqrt{-E} \rho)$ are 
expressed in terms of the gamma function and the modified Bessel function. 
In the limit $b \to \infty$, the bound state goes to the threshold, where it 
turns to the virtual state, which persists for $b < 0$ and whose energy is 
given by the above expression. 
The above expressions for $|a| \to \infty$ are a good approximation for 
the properties of the deep state, which exists for $|a|/b \gg 1$. 
Note also that redefinition of the parameter $\tilde{b} = 
\frac{b}{2}\left[-\Gamma(-\gamma)/\Gamma(\gamma)\right]^{\frac{1}{2\gamma}}$ 
gives the usual relation, $E = -\tilde{b}^{-2}$, between the energy and 
the scattering length. 


To elucidate the qualitative features of the problem in connection with 
the three-body boundary condition, one constructs a simple model that 
provides reliable dependence of the bound-state energy on $b$ and $m/m_1$. 
The model is based on splitting the Hamiltonian into two parts: 
the singular one containing terms singular as $\rho \to 0$ and the remaining 
one describing a smooth dependence on $m/m_1$. 
The former part is defined as one HRE of~(\ref{system1}) containing 
the smallest $\gamma_n^2(\rho)$, moreover, only singular terms 
$(\gamma^2 - 1/4)/\rho^2 + q/\rho$ are retained, which allows one to obtain 
the correct behaviour of the solution for $\rho \to 0$ and to reproduce 
the attraction for finite $\rho$. 
The remaining part is defined simply as a constant $\epsilon(m/m_1)$. 
Explicitly, one comes to the equation 
$\displaystyle \left(\frac{d^2}{d \rho^2} - \frac{\gamma^2 - 1/4}{\rho^2} - 
\frac{q}{\rho} + E - \epsilon \right)f(\rho) = 0 $, 
whose square-integrable solution is written as 
$f(\rho) = \rho^{1/2 + \gamma } e^{-\kappa \rho} 
\Psi \left(1/2 + \gamma + q/(2 \kappa ), 1 + 2 \gamma ; 
2 \kappa \rho \right)\ $, 
where $\kappa = \sqrt{\epsilon - E}$ and $\Psi (a, c; z)$ is the confluent 
hyper-geometric function decaying as $z \to \infty$. 
The eigenenergy equation 
\begin{equation}
\label{traneq_simple_model}
\left(2\kappa |b|\right)^{2\gamma } = 
\displaystyle \mp \frac{\Gamma (2 \gamma ) 
\Gamma \left(1/2 - \gamma + q/(2 \kappa) \right)} 
{\Gamma (-2 \gamma ) \Gamma \left(1/2 + \gamma + q/(2 \kappa) \right)} \  
\end{equation} 
follows from boundary condition~(\ref{as_gam121}) for all $0 < \gamma < 1$ 
($\mu_c > m/m_1 > \mu_r$) except $\gamma = 1/2$ ($m/m_1 = \mu_e$). 
The eigenenergy equation for $\gamma = 0$ is obtained either by taking 
the limit in Eq~(\ref{traneq_simple_model}) or from the boundary 
condition~(\ref{as_gam0}) that gives $\log (2 \kappa b) + 
\psi \left(\frac{1}{2} + \frac{q}{2\kappa} \right) + 2\gamma_C  = 0$ for 
$b > 0$. 
Hereafter, $\psi(x)$ is the digamma function and $\gamma_{C} \approx 0.5772$ 
is the Euler--Mascheroni constant.
In the special case of $\gamma = 1/2$ ($m/m_1 = \mu_e$) the eigenenergy 
equation $\frac{1}{q}\left( \frac{1}{ b} - \kappa\right) - 
\log \left( \frac{|q|}{2 \kappa}\right)+ 
\psi \left(1 + \frac{q}{2 \kappa}\right) + 
2 \gamma_{C} - 1 = 0$ comes from~(\ref{as_gam12}). 

\begin{figure*}[hptb]
\hspace{-.5cm}
\includegraphics[width=0.5\textwidth]{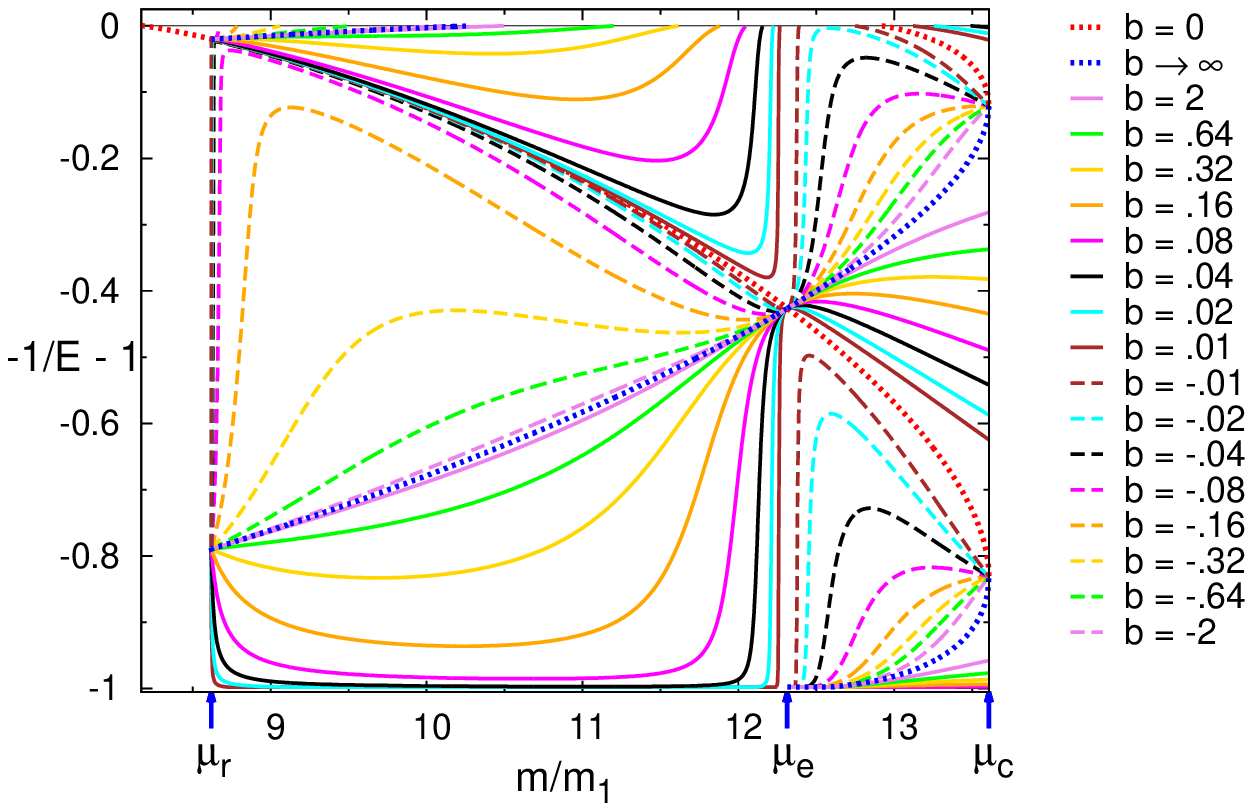} 
\includegraphics[width=0.5\textwidth]{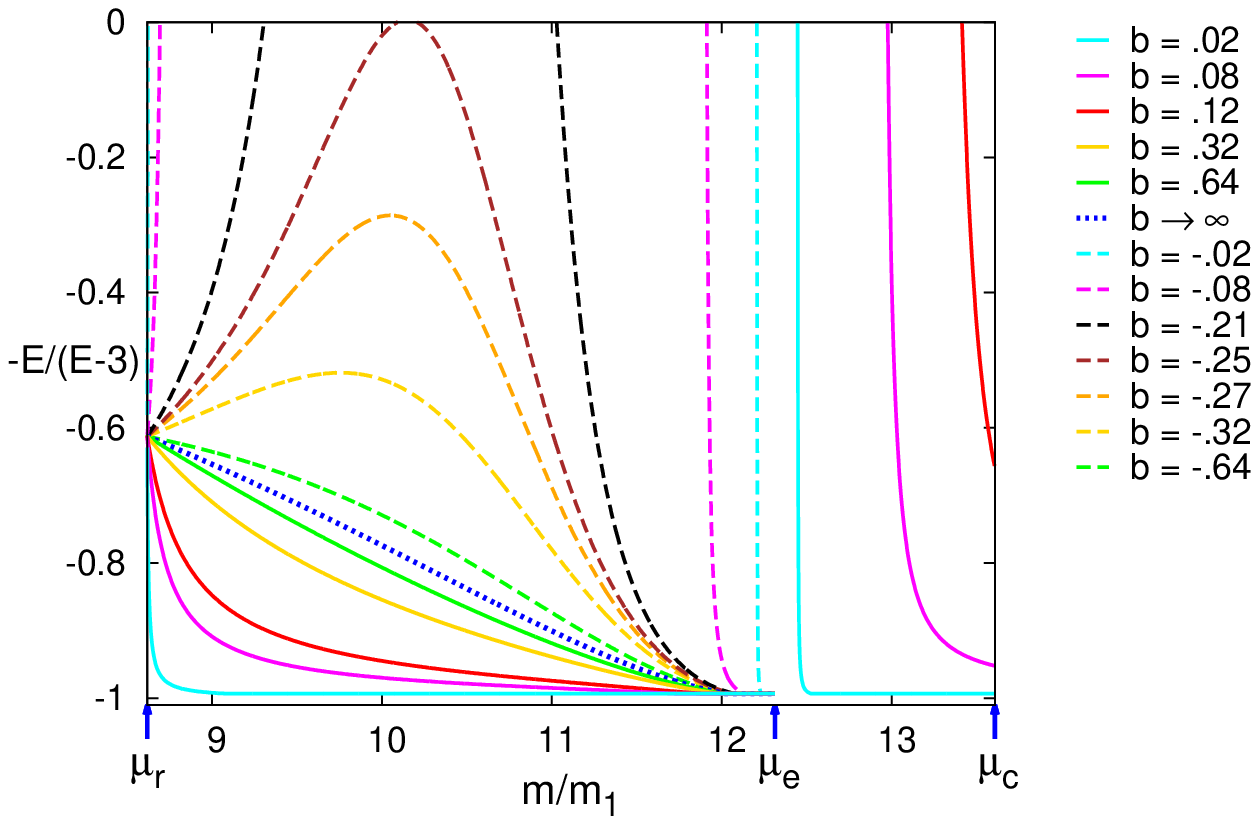}
{\caption{Bound-state energies $E$ as a function of $m/m_1$ and $b$ for 
the two-body scattering length $a > 0$ (left) and $a < 0$ (right) and 
the energy axis scaled to map $-\infty < E < -1$ (left) and 
$-\infty < E < 0$ (right) to the interval $(-1, 0)$. 
Values $\mu_r$, $\mu_e$ and $\mu_c$ correspond to $\gamma = 1$, $1/2$ and $0$.
} 
\label{fig_en}}
\end{figure*}

The simple model is equivalent to the generalised Coulomb problem 
incorporating the zero-range interaction. 
As follows from Eq.~(\ref{traneq_simple_model}), the bound-state energies 
monotonically increase with increasing $b$; moreover, one bound state 
appears if $b$ passes through zero. 
It is helpful to examine two limiting cases of $b = 0$ and $b \to \infty$, 
which gives the eigenvalues 
$\kappa_{nb} = -\frac{q}{2 (n + s_b \gamma ) + 1}$, where $n$ is a non-negative 
integer and $s_0 = +1$ ($s_\infty = -1)$. 
The bound-state energies are 
\begin{equation}
\label{Enb0inf}
E_{nb} = -\frac{q^2}{ [2 (n + s_b \gamma ) + 1]^2} + \epsilon \ , 
\end{equation}
where $n$ is restricted by the condition $2 (n + s_b \gamma ) + 1 > 0$ if 
$a > 0$ ($q < 0$) and $2 (n + s_b \gamma ) + 1 < 0$ if $a < 0$ ($q > 0$). 
Hereafter it is convenient to take $|a|$ as a length unit that sets
the two-body binding energy to unity. 
Estimating the constant $\epsilon \approx -0.5$, one finds that for $a > 0$ 
there are two branches below the threshold (at $E \le -1$) if $b = 0$ and 
three branches if $b \to \infty$, while for $a < 0$ there is one branch below 
the threshold (at $E \le 0$) if $b \to \infty$ (see Fig.~\ref{fig_en}). 
For $a > 0$, from Eq.~(\ref{Enb0inf}) follows degeneracy of the branches 
$E_{n0}$ and $E_{n\infty}$ ($n = 0, 1$) as $m/m_1 \to \mu_c$ ($\gamma \to 0$), 
$E_{00}$ and $E_{1\infty}$ as $m/m_1 \to \mu_e$ ($\gamma \to 1/2$), and $E_{00}$ 
and $E_{2\infty}$ as $m/m_1 \to \mu_r$ ($\gamma \to 1$). 
Moreover, from Eq.~(\ref{traneq_simple_model}) it follows that as 
$m/m_1 \to \mu_c$ ($\gamma \to 0$) the energies for any $b < 0$ converge 
to either $E_{00} = E_{0\infty}$ or $E_{10} = E_{1\infty}$. 
As $m/m_1 \to \mu_e$ ($\gamma \to 1/2$) the energies converge to either of 
three options, the threshold $E = -1$, $E_{00} = E_{1\infty}$, and $-\infty$. 
And as $m/m_1 \to \mu_r$ ($\gamma \to 1$) the energies converge to either 
$E_{1\infty}$ or $E_{00} = E_{2\infty}$ as shown in Fig.~\ref{fig_en}. 
For $a < 0$, the energies converge to $E_{0\infty}$ as $m/m_1 \to \mu_r$ 
($\gamma \to 1$) and to $-\infty$ as $m/m_1 \to \mu_e$ ($\gamma \to 1/2$) 
for $b \ne 0$. 


The three-body bound-state energies are determined by numerical solution of 
the truncated system of HRE~(\ref{system1}) complemented by boundary 
conditions~(\ref{as_gam121}),~(\ref{as_gam0}), and~(\ref{as_gam12}). 
The numerical method is the same as in~\cite{Kartavtsev07,Kartavtsev07a} apart 
from implementation of the boundary conditions at sufficiently small $\rho$. 
Sufficient accuracy of the calculated three-body bound-state energies is 
achieved by solving up to eight HREs; the results are plotted in 
Fig.~\ref{fig_en}. 
The calculated dependences are consistent with the overall predictions of 
the simple model. 
The energy dependence on $b$ for fixed $m/m_1$ is typical of a sum of 
the finite-range and zero-range potentials, in particular, variation of 
the parameter $b$ leads to the appearance or disappearance of one bound state. 

The calculations for $a > 0$ show that if $m/m_1 \to \mu_r$ the energies for 
any $b$ converge either to $E_{1\infty} \sim -4.7477$ or to 
$E_{00} = E_{2\infty} \sim -1.02090$, if $m/m_1 \to \mu_e$ there is one limit 
$E_{00} = E_{1\infty} \sim -1.74397$, and if $m/m_1 \to \mu_c$ the energies for 
any $b \le 0$ converge either to $E_{00} = E_{0\infty} \to -5.89543$ or 
to $E_{10} = E_{1\infty} \to -1.13767$. 
In agreement with~\cite{Kartavtsev07} it is found that if $m/m_1 \le \mu_r$, 
where only $b = 0$ is allowed, there is one bound state, which arises at 
$m/m_1\approx 8.17259$ and naturally continues the branch $E_{00}$. 
The calculations for $a < 0$ show that if $m/m_1 \to \mu_r$ the energies for 
any $b$ converge to the limit $E_{0\infty} \to -4.7147$. 
If $m/m_1 \to \mu_r$, the limit $E_{0\infty}$ for $a < 0$ coincides with 
the limit $E_{1\infty}$ for $a > 0$, as predicted by the simple 
model~(\ref{Enb0inf}). 

\begin{figure*}[hptb]
\includegraphics[width=.36\textwidth]{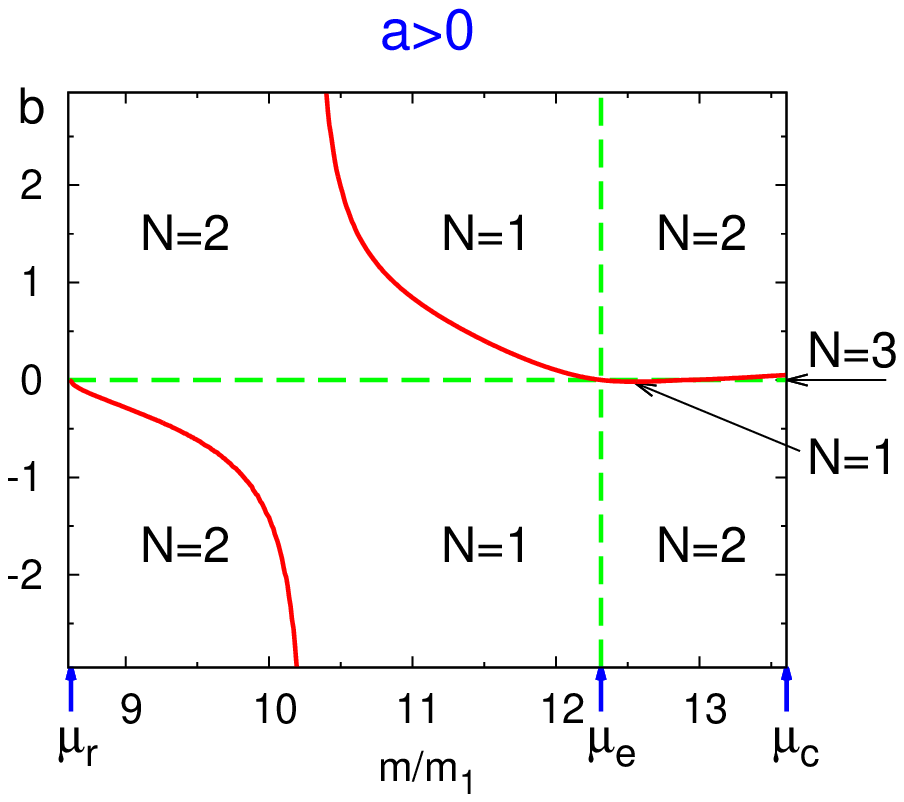} \hspace{-1.cm} 
\includegraphics[width=.36\textwidth]{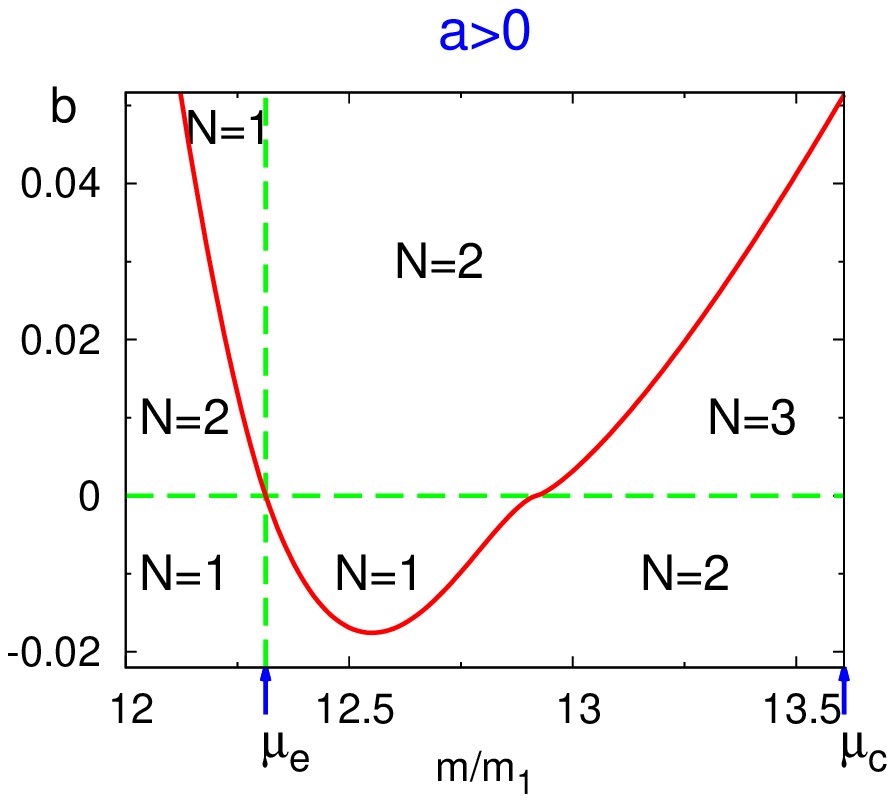}\hspace{-1.cm}
\includegraphics[width=.36\textwidth]{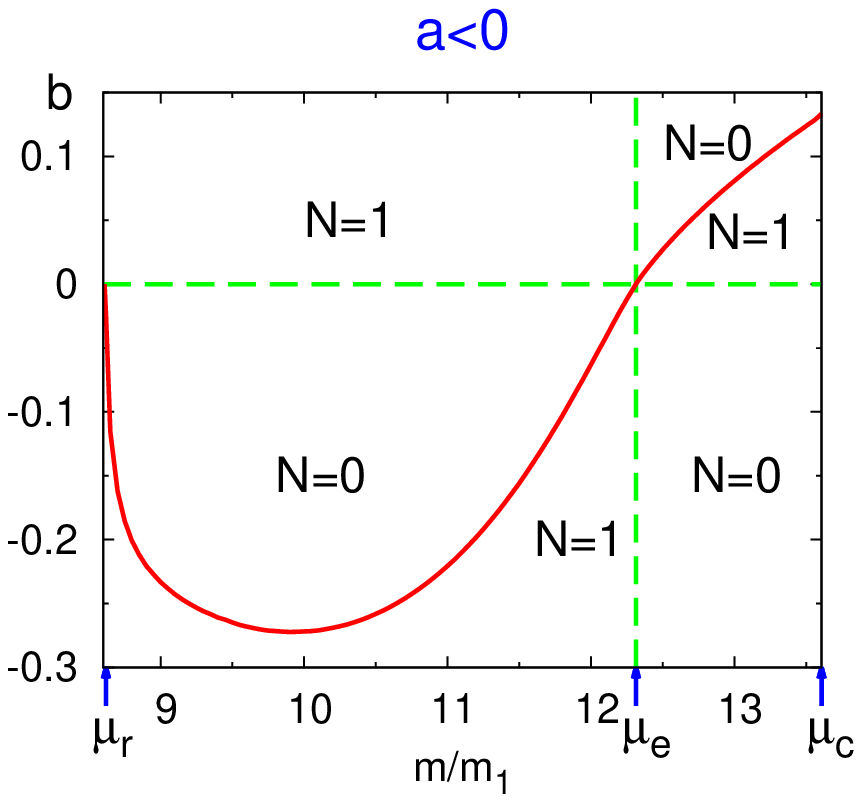}
{\caption{
A number of bound states in each domain of the $m/m_1$ -- $b$ plane. 
Solid (red) line: critical three-body parameter $b_c(m/m_1)$ corresponding to 
the bound-state energy at the threshold. 
Dashed (green) lines: domain boundaries determined by $m/m_1 = \mu_e$ and 
$b = 0$. 
Part of the left panel is plotted in the middle panel to discern details. 
Values $\mu_r$, $\mu_e$ and $\mu_c$ correspond to $\gamma = 1$, $1/2$ and $0$. 
} 
\label{fig_mub}}
\end{figure*}

Elaborate calculations were carried out to determine the critical parameter 
$b_c(m/m_1)$, for which the bound-state energy coincides with 
the threshold~\cite{footnote1}. 
The lines $b_c(m/m_1)$, $b = 0$, and $m/m_1 = \mu_e$ form boundaries
of the domains of the definite number of bound states in the $m/m_1$ -- $b$ 
plane as presented in Fig.~\ref{fig_mub}. 
Few points of the dependence $b_c(m/m_1)$ are of special interest,~viz., one 
finds for $a > 0$ that $b_c = 0$ at $m/m_1 \approx 12.91742$, 
$b_c \to \pm \infty$ at $m/m_1 \approx 10.2948$, $b_c \approx 0.05166$ at 
$m/m_1 = \mu_c$, and $b_c(m/m_1)$ has a local minimum $b_c \approx -0.01754$ 
at $m/m_1 \approx 12.550$. 
Similarly, one finds for $a < 0$ that $b_c \approx 0.13620$ at $m/m_1 = \mu_c$, 
and $b_c(m/m_1)$ has a local minimum $b_c \approx -0.2501$ at 
$m/m_1 \approx 10.15$. 


Until now, in a number of reliable investigations of three two-component 
fermions (for $m/m_1 \le 
\mu_c$)~\cite{Petrov03,Kartavtsev07,Kartavtsev07a,Endo11,Helfrich11} it was 
explicitly or implicitly assumed that only one particular form of the wave 
function near the triple-collision point is allowed, which in terms of this 
Letter means that the three-body parameter $b$ was set to zero. 
Nonetheless, the problem of two linear-independent square-integrable solutions 
was mentioned in~\cite{Petrov03,Werner06a,Nishida08,Safavi-Naini13}. 
The two-parameter variety of three-body problems was defined 
in~\cite{Safavi-Naini13} by introducing the logarithmic derivative of the wave 
function at small hyper-radius; the relation of the present results and those 
of~\cite{Safavi-Naini13} is discussed in~\cite{footnote1}.   
A rigorous treatment of few two-component fermions with the contact two-body 
interactions and the construction of a self-adjoint Hamiltonian was discussed 
from the mathematical point of view 
in~\cite{Minlos12,Minlos14a,Correggi12,Correggi15}. 
The approach of~\cite{Correggi12} was further exploited in the calculation 
of three-body bound states~\cite{Michelangeli13}. 

The transition from the infinite Efimov spectrum to the one-parameter spectrum 
described in this Letter under the variation of the mass ratio is a general 
scenario, which will appear in a number of problems. 
One should anticipate the same transition for any problem, whose essential 
properties are determined by the effective potential with the singular part 
$\sim x^{-2}$, if its strength depends on a parameter (similar to the mass 
ratio). 
Evident example of this kind is the problem of three two-species particles 
in any $L^P$ sectors~\cite{Kartavtsev07a,Endo11,Helfrich11}. 
Similar to the case of the $L^P = 1^-$ sector, the three-body parameter should 
be introduced in the $L^-$ sectors of odd $L$ and in the $L^+$ sectors of even 
$L > 0$ if two identical particles are fermions and bosons, respectively. 
Also, this scenario will be realised for the three-body problem in the mixed 
dimensions~\cite{Nishida08a,Nishida11} or in the presence of spin-orbit 
interaction~\cite{Shi14,Shi15}. 

In future studies it is natural to find $m/m_1$ and $b$ dependences of 
the scattering cross sections, three-body resonances, and recombination rates. 
The disclosed dependence on the three-body parameter should be taken into 
account in many-body properties as well; promising examples are the four-body 
($3 + 1$)~\cite{Castin10} and ($2 + 2$)~\cite{Endo15a} problems.
Furthermore, the three-body parameter will be important in the crossover 
problem~\cite{Endo12}, i.~e., in the relation of solutions for $m/m_1$ below 
and above $\mu_c$; another interesting point is the crossover of 
the solutions for $m/m_1$ above and below $\mu_r$. 

\begin{acknowledgments}

OIK acknowledges support from the Institute for Nuclear Theory during 
the program {\em Universality in Few-Body Systems: Theoretical Challenges 
and New Directions, INT-14-1.} 

\end{acknowledgments}

\bibliography{fermions,ferm_zrm,ferm_footnote}

\clearpage

\appendix*
\section*{Supplemental Material: Universal description of three 
two-component fermions}

\renewcommand{\theequation}{S\arabic{equation}}
\renewcommand{\thefigure}{S\arabic{figure}}
\setcounter{equation}{0}
\setcounter{figure}{0}

\subsection*{Mass-ratio dependences of $\gamma$ and $q$}

\begin{figure}[hbt]
\includegraphics[width=.48\textwidth]{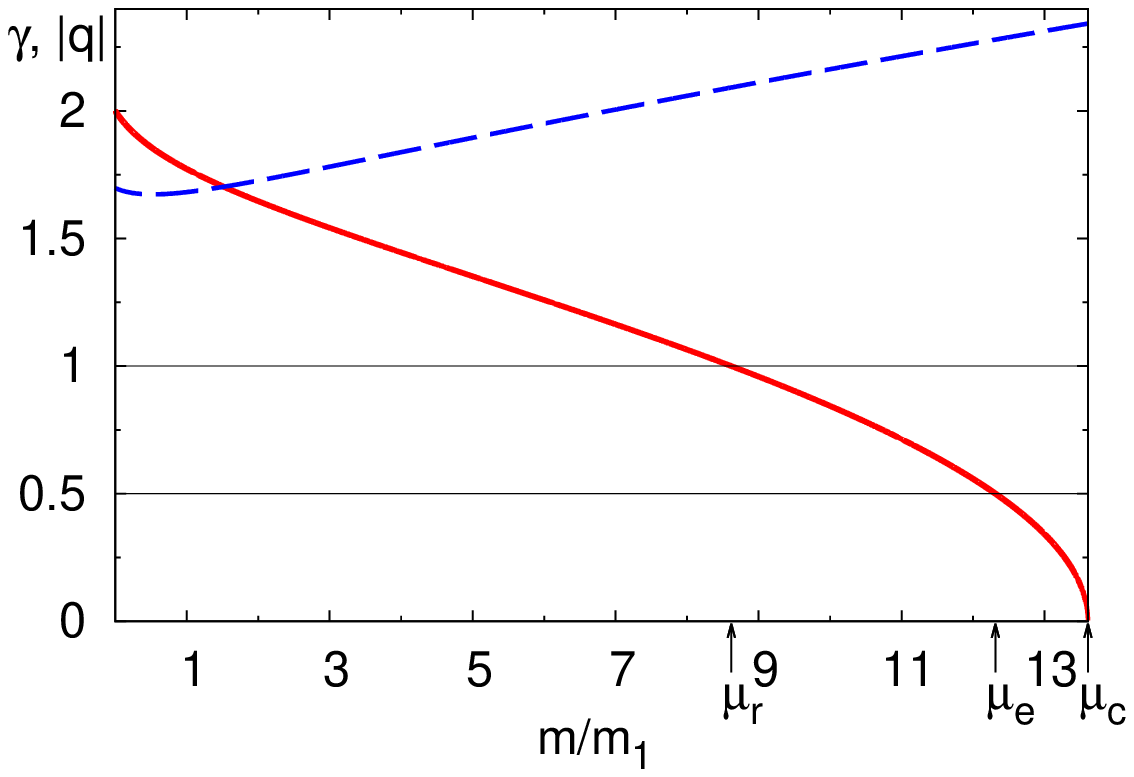}
{\caption{The dependences of $\gamma$ and $|q|$ on $m/m_1$ are depicted 
by the solid (red) and dashed (blue) lines, respectively. 
Values $\mu_r$,  $\mu_e$, and $\mu_c$ correspond to $\gamma = 1$, $1/2$, and 
$0$.} 
\label{fig_gammaq}}
\end{figure}

The smallest eigenvalue of the auxiliary problem on a hyper-sphere 
$\gamma^2 \equiv \gamma^2(0)$ and its derivative 
$q = \displaystyle\left[\frac{d\gamma^2(\rho)}{d\rho}\right]_{\rho = 0}$ 
are shown in Fig.~\ref{fig_gammaq}. 
Note that the two-body scattering length is taken as a length unit ($|a| = 1$) 
and $q < 0$ ($q > 0$) for $a > 0$ ($a < 0$). 

\subsection*{Special mass-ratio values $\mu_r$,  $\mu_e$, and $\mu_c$ }

Few values of the mass ratio are of special interest, namely, $\mu_r$, 
$\mu_e$, and $\mu_c$ correspond to $\gamma = 1$, $1/2$, and $0$.
Using Eq.~(\ref{transeq}) in the limit $\rho \to 0$ one comes to the equations 
\begin{equation}
 (\sin\omega_r + 1/2) \sin 2 \omega_r - \omega_r = 0\ , 
\end{equation}
\begin{equation}
 \cos\omega_e \cos\frac{\omega_e}{2} - \frac{\sqrt{2}}{3} 
\tan\frac{\omega_e}{2} = 0\ , 
\end{equation}
\begin{equation}
\frac{\pi}{2} \sin^2\omega_c - \tan\omega_c + \omega_c = 0\ . 
\end{equation}
Recall the definition $\sin\omega = 1/(1 + m_1/m)$. 
The roots of these equations are $\omega_r \approx 1.11075583$, 
$\omega_e \approx 1.18073571$, and $\omega_c \approx 1.19862376$ that 
correspond to the mass-ratio values $\mu_r \approx 8.61857692$, 
$\mu_e \approx 12.3130993$, and $\mu_c \approx 13.6069657$. \\ 

\subsection*{Three-body boundary conditions}

It is suitable to write the three-body boundary conditions in the alternative 
form,~viz., in terms of the derivative of the channel function $f(\rho)$. 
The boundary condition for $\mu_r < m/m_1 < \mu_c$ ($1 > \gamma > 0$), 
except $m/m_1 = \mu_e $ ($\gamma = 1/2$), which is equivalent to 
Eq.~(\ref{as_gam121}), reads 
\begin{equation} 
\label{bc_gam121}
\displaystyle\lim_{\rho \to 0} \left( \rho^{1 - 2\gamma } 
\frac{d}{d\rho } \pm \frac{2\gamma}{|b|^{2\gamma}}\right) 
\frac{\rho^{\gamma - 1/2}} {1 - 2\gamma + q\rho}  
f(\rho) = 0 \ .
\end{equation} 
In the limit $m/m_1 \to \mu_c$ ($\gamma \to 0$) the boundary condition, 
which is equivalent to Eq.~(\ref{as_gam0}), takes the form 
\begin{equation} 
\label{bc_gam0}
\displaystyle \lim_{\rho \to 0} \left( \rho \frac{d}{d\rho } - 
\frac{1}{\log (\rho /b)}\right) \rho^{-1/2}f(\rho) = 0\ ,
\end{equation}
where only $b > 0$ is allowed. 
In the specific case of $\gamma = 1/2$ ($m/m_1 = \mu_e$) the boundary 
condition 
\begin{equation} 
\label{bc_gam12}
\displaystyle\lim_{\rho \to 0} \left( \frac{d}{d\rho} + \frac{1}{b}\right) 
\frac{f(\rho)}{1 + q\rho\log \rho} = 0\ 
\end{equation} 
is equivalent to Eq.~(\ref{as_gam12}). 
Notice that the boundary condition for $\gamma = 0$ determined by 
Eq.~(\ref{as_gam0}) or Eq.~(\ref{bc_gam0}) is similar to that for the $2D$ 
zero-range model~\cite{Kartavtsev06}, whereas for $\gamma = 1/2$ the boundary 
condition of the form~(\ref{as_gam12}) or~(\ref{bc_gam12}) is similar to that 
for a sum of the zero-range and Coulomb potentials~\cite{Yakovlev13}. 

\subsection*{Solution for $m/m_1 = \mu_e$}

\begin{figure*}[htb]
\includegraphics[width=.45\textwidth]{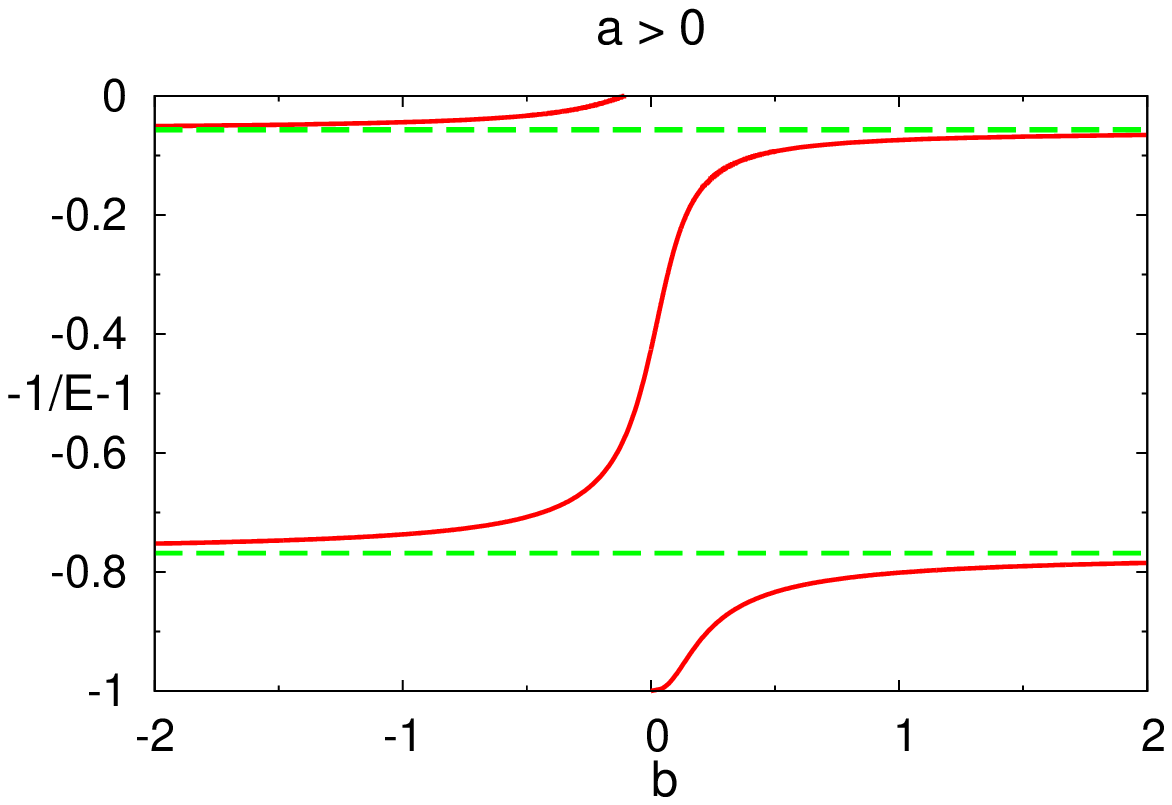}
\includegraphics[width=.45\textwidth]{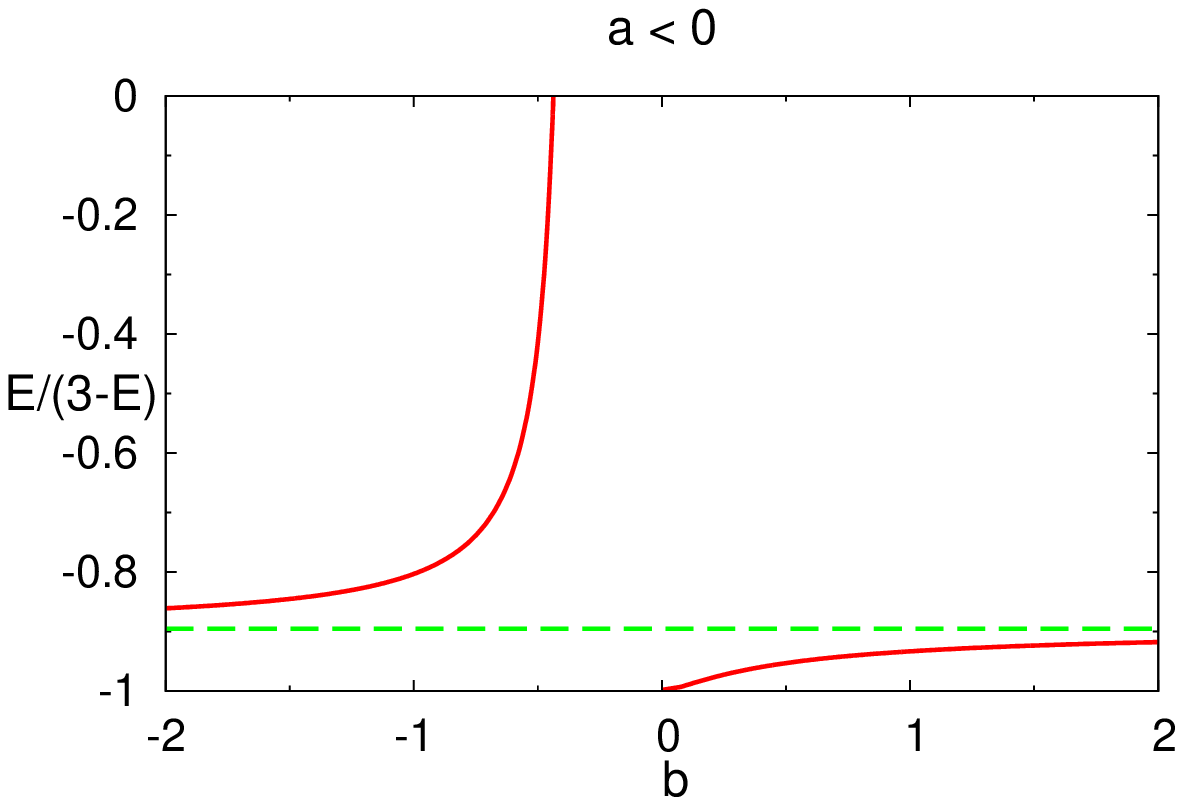}
{\caption{Bound-state energies $E$ as a function of $b$ at $m/m_1 = \mu_e$ are 
plotted by solid (red) lines and asymptotic limits for $b \to \infty$ are 
indicated by dashed (green) lines. 
The two-body scattering length $a > 0$ (left) and $a < 0$ (right) and 
the energy axis scaled to map $-\infty < E < -1$ (left) and $-\infty < E < 0$ 
(right) to the interval $(-1, 0)$. 
} 
\label{Fig_mu_e} }
\end{figure*}

A noticeable feature of the problem near $m/m_1 = \mu_e$ ($\gamma = 1/2$) is 
the degeneracy of energy dependences for different $b$ and a lack of 
continuity in the definition of $b$. 
It is not surprising as the sign of the most singular term in HRE alters 
if $\gamma $ goes across $1/2$. 
Due to discontinuity in the definition of $b$ the limiting values of 
the bound-state energy for $m/m_1 \to \mu_e \mp 0$ ($\gamma \to 1/2 \pm 0$) do 
not coincide with each other and with that calculated exactly at 
$m/m_1 = \mu_e$ ($\gamma = 1/2$). 
The dependence of the bound-state energy on $b$ is calculated using 
boundary condition~(\ref{as_gam12}) and plotted in Fig.~\ref{Fig_mu_e}. 
Notice that in boundary condition~(\ref{as_gam12}) one could substitute 
$\log \rho$ with $\log (\rho/\rho_0)$ introducing a scale $\rho_0$, which 
simply leads to redefinition of length $\tilde {b} = b/(1 - b \log \rho_0)$. 

The calculations for $a > 0$ show that there are two bound states, one of 
which disappears for $-.108 < b \le 0$; for $a < 0$ there is one bound state, 
which disappears for $-.437 < b \le 0$. 
In the limit $b \to \infty$ the bound-state energies tend to $-4.319$ and 
$-1.061$ for $a > 0$ and to $-25.720 $ for $a < 0$. 
For $b = 0$ definitions~(\ref{as_gam121}) and~(\ref{as_gam12}) are the same 
and for $a > 0$ the bound-state energy takes the value $E_{00} \sim -1.74397$. 

\subsection*{Zero-range limit of the three-body potential}

Consider simple examples of the transition to the zero-range limit to clarify 
the introduction of the boundary condition at the triple-collision point 
$\rho \to 0$. 

\subsubsection*{Square-well potential}

Find the connection of the three-body parameter $b$ and the parameters of 
the regularised potential defined as the square well $U(\rho) = -U_0$ for 
$\rho \le \rho_0$ and as 
$U(\rho) = \frac{\gamma^2 - 1/4}{\rho^2} + \frac{q}{\rho} $
for $\rho > \rho_0$ in the zero-range limit $\rho_0 \to 0$. 
The solution of the equation 
$\displaystyle \left(\frac{d^2}{d \rho^2} - U(\rho) + E \right)f(\rho) = 0 $ 
is $f(\rho ) = \cos \kappa \rho $ ($\kappa = \sqrt{U_0 + E}$) for 
$\rho \le \rho_0$ and is of the form~(\ref{as_gam121}) for $\rho > \rho_0$, 
which gives the relation 
\begin{equation}
\label{kaprho}
\kappa\rho_0\tan\kappa\rho_0 = \gamma - \frac{1}{2} \pm 
2 \gamma \left( \frac{\rho_0}{|b|}\right)^{2 \gamma } - 
\frac{q \rho_0}{1 - 2 \gamma + q \rho_0} \ . 
\end{equation} 
Up to the leading-order terms containing $b$ and $q$, the potential strength 
$U_0$ is related to the interaction range $\rho_0$ as 
\begin{widetext}
\begin{equation}
\label{potrho}
U_0 = v \left[ \frac{1}{\rho_0^2} \pm \frac{4\gamma}
{|b|^{2\gamma}(\gamma^2 - 1/4 + v)\rho_0^{2(1 - \gamma )}} + 
\frac{q}{(\gamma^2 - 1/4 + v)(\gamma - 1/2)\rho_0} \right] \ , 
\end{equation} 
\end{widetext}
where $v$ is determined by $\sqrt{v}\tan {\sqrt{v}} = \gamma - 1/2$. 
Thus, the most singular term $\sim \rho_0^{-2}$ in the dependence 
$U_0(\rho_0)$ is determined by $\gamma $, whereas the parameter $b$ determines 
less singular terms. 
With decreasing $\gamma $, the higher order terms containing $b$ prevail over 
the term proportional to $q$, e.~g., for $\gamma < 1/4$, the higher order 
term $\sim \rho_0^{-2 + 4 \gamma }$ is more important than that of $q/\rho_0$. 
For $\gamma = 0$ Eqs.~(\ref{kaprho}) and~(\ref{potrho}) take the following 
form: 
$\displaystyle\kappa \rho_0 \tan \kappa \rho_0 = 
-\frac{1}{2} - \frac{1}{\log (\rho_0/b)} $ 
and $ U_0 = \displaystyle\frac{v}{\rho_0^{2}} \left[ 1 + \frac{2}
{(1/4 - v)\log (\rho_0/b)} \right] $. 
If $b = 0$, relation~(\ref{kaprho}) is not applicable; in this case 
the form~(\ref{as_gam121}) gives 
$\displaystyle\kappa \rho_0 \tan \kappa \rho_0 + \gamma + \frac{1}{2} = 0$ and 
$ U_0 = \displaystyle\frac{\tilde v}{\rho_0^{2}}$, where 
$\sqrt{\tilde v}\tan {\sqrt{\tilde v}} = -\gamma - 1/2$. \\

\subsubsection*{Set-up of the logarithmic derivative}

The wave function in the vicinity of the triple-collision point can be 
specified by imposing the three-body boundary condition for small $\rho_0$, 
e.~g., by setting the dimensionless logarithmic derivative of the channel 
function $\tan\delta = 
\rho \displaystyle\frac{d \log f}{d \rho }$~\cite{Safavi-Naini13}. 
Using the asymptotic form of the solution as $\rho \to 0$~(\ref{as_gam121}), 
one readily finds that for $\rho_0 \to 0$ two parameters $\delta$ and 
$\rho_0$ are related to the three-body parameter $b$ as
\begin{widetext}
\begin{equation}
\label{b_blume}
|b|^{2\gamma} = \pm \rho_0^{2\gamma}
\frac{\tan\delta - \gamma - 1/2}{\left[1 + q \rho_0/(1 - 2\gamma)\right] 
\tan\delta + \gamma - 1/2 + q \rho_0(\gamma - 3/2)/(1 - 2\gamma) } \ , 
\end{equation} 
\end{widetext}
except for $\gamma = 1/2$. 
This relation could be used to link the results of~\cite{Safavi-Naini13} 
and those of the present Letter. 
For example, the dependence of the bound-state energy on $\delta$ 
in~\cite{Safavi-Naini13} is discontinuous at some $\delta_{cr}$ depending on 
$\rho_0$. 
It stems from the discontinuous dependence of the parameter $b$ on $\delta$ 
and $\rho_0$, 
\begin{equation}
\label{tandc}
\tan\delta_{cr} = \frac{(1 - 2\gamma )^2 + q \rho_0 (3 - 2\gamma )}
{2(1 - 2\gamma + q \rho_0)} \ , 
\end{equation} 
which follows from~(\ref{b_blume}) and for $\rho_0 \to 0$ takes the form 
$\tan\delta_{cr} = 1/2 - \gamma$, excluding the neighbourhood $\sim q \rho_0$ 
of the point $m/m_1 = \mu_e$ (of the order of $|\gamma - 1/2| < q \rho_0$). 
This exact expression can be compared with the dependence $\delta_{cr}(m/m_1)$, 
which was numerically calculated and presented in Fig.~5 
of~\cite{Safavi-Naini13}. 
In particular, the exact expression gives that
$\delta_{cr} \to \arctan(1/2) \approx 0.46$ for $m/m_1 \to \mu_c$; 
the discrepancy with $\delta_{cr}$ in Fig.~5 of Ref.~\cite{Safavi-Naini13} 
indicates difficulty of the calculation in this mass-ratio limit. 

\subsection*{Threshold solution}

Critical dependence $b_c(m/m_1)$ for the appearance or disappearance of 
the bound state is determined by solving the eigenvalue problem for HREs 
at the two-body threshold $E = -1$ for the two-body scattering length $a > 0$ 
and at the three-body threshold $E = 0$ for $a < 0$. 
The square-integrability of solution follows from asymptotic behaviour of 
the HRE $\displaystyle\left[\frac{d^2}{d \rho^2} - U_{eff}(\rho) + 
E \right] f(\rho) =  0$, where $U_{eff} \to -1 + 2/\rho^2$ for $a > 0$ and 
$U_{eff} \to 35/(4 \rho^{2})$ for $a < 0$ as 
$\rho \to \infty$~\cite{Kartavtsev07}. 
Thus, if $\rho \to \infty$ the channel function $f(\rho)$ decays as 
$\rho^{-1}$ for $a > 0$ and as $\rho^{-7/2}$ for $a < 0$. 
As usual, the bound state at the threshold turns to a narrow resonance 
under small variations of $m/m_1$ and $b$. 

\end{document}